# Catalysis for *e*-chemistry: need and gaps for a future de-fossilized chemical production, with focus on the role of complex (direct) syntheses by electrocatalysis

*Georgia PAPANIKOLAOU[†], Gabriele CENTI*[†], Siglinda PERATHONER*[†] and Paola LANZAFAME[†]*

[†] University of Messina, Dept. ChiBioFarAm, ERIC aisbl and CASPE/INSTM, V. le F. Stagno d'Alcontres 31, 98166 Messina, Italy.

* Corresponding authors: GC - centi@unime.it: SP - perathon@unime.it

KEYWORDS:  e-chemistry; de-fossilized chemical production; reactive catalysis; fossil fuels beyond; renewable energy; petrochemistry

GRAPHICAL ABSTRACT          (8,5 x 3,2 cm)

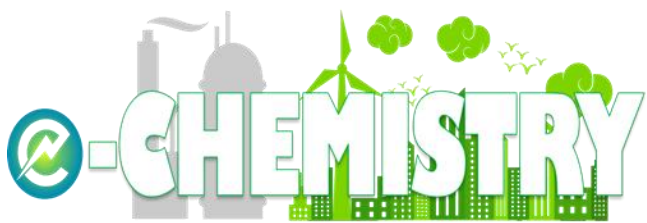

ABSTRACT

The prospects, needs and limits in current approaches in catalysis to accelerate the transition to an *e*-chemistry, where this term indicates a novel fossil fuel-free chemical production, are discussed. It is suggested that an *e*-chemistry is a necessary element of the transformation to meet the targets of net zero emissions by year 2050 and that this conversion from the current petrochemistry is feasible. However, the acceleration of the development of catalytic technologies based on the use of renewable energy sources (indicated as reactive catalysis) is necessary, evidencing that these are part of a system changes and thus should be assessed from this perspective. However, it is remarked that the current studies in the area are not properly addressing the needs to develop the catalytic technologies required for *e*-chemistry, presenting a series of relevant aspects and directions on which research should be focused to develop the framework system transformation necessary to implement an *e*-chemistry.

# ■ INTRODUCTION

The research interest on catalysis based on the direct use of renewable energy sources - RES (photo-, electro- and plasma-catalysis as the main methodologies) is fast rising.[1-10] As index of the growing of activities, the papers with electrocatalysis as keyword raised from about 30 in year 2000 to nearly 600 in 2021, and those containing the term photocatalysis from about 40 to nearly 1000 per year in the same period. However, most of the research activity focused on only few molecules ($CO_2$, $N_2$, $H_2O$, and few biobased chemicals). On the other hand, meeting the targets fixed at political level (for example, reach a net zero greenhouse gas emission target by year 2050, as committed by EU) would require a more disruptive effort than developing some catalytic processes driven directly from RES.[10] It is necessary to consider the possibility to substitute largely, if not entirely, the use of fossil fuels (FF) both as energy and carbon source, the latter particularly for chemical production. The concept of *e*-chemistry indicates that this (almost) FF-free chemical production, which can be identified as part of the overall transformation necessary, can meet the targets of net zero emissions (NZE) for year 2050 and beyond. Implementing *e*-chemistry (and associated *e*-refinery) concept will address some important societal challenges which are facing to realize a resilient and sustainable future: i) overcome intermittency of RES, ii) implement a world-economy based on the distribution and long-distance transport of renewable energy, iii) develop $CO_2$ neutral or even carbon-negative technologies to supply the goods and energy necessary for the society, and iv) realize a de-fossilized chemical industry. *e*-Chemistry is particularly associated to the last objective but is also closely related to the others.

To realize an *e*-chemistry requires a combination of actions, as

- the direct electrification of the operations in chemical industry using FF as energy input (for example, most of the furnaces in chemical processes use FF to provide the heat necessary

for operations)[11-14]

- the closure of the carbon cycle in chemical production by introducing novel technologies for the molecular reuse of waste and end-of-life chemicals (from $CO_2$ to end-of-life chemicals as recycled plastics)[15-17]

- the realization of novel chemical processes using RES as energy input for the process, where electro-, photo- and plasma-catalysis represent the main technological options.[8,18-21]

Catalysis plays a crucial role in several of these novel technologies allowing not only to develop innovative routes with a large decrease in the carbon footprint, up to over 90%, but also realize process intensification with thus reduction of fixed costs and better suitability to a distributed model of chemical production.[8,22-24] This combination of drastic reduction in the carbon footprint, use of alternative raw materials to FF and process intensification (with associated impacts such as the possibility to develop distributed production modes as well as faster and more flexible industrialization by parallel units) is the reading key to analyze the potential of catalytic technologies based on the use of RES, rather than limiting to purely economic considerations.

However, addressing this challenge for catalysis requires also to reconsider the fundamental bases of catalysis science and technology. We introduced the term *reactive* catalysis to differentiate photo-, electro- and plasma-catalysis from the conventional *thermal* catalysis.[10] In the latter case, energy in the form of heat is provided to overcome the activation energy, while in reactive catalysis already highly energetic species (electron, holes, radicals, vibrationally-excited species) are generated by application of an electrical potential, by light irradiation or by generation of a non-thermal plasma. The fundamental modes of operations, and consequently the design aspects for the catalysts are different. It is thus necessary to use novel methodologies to understand and develop reactive catalysis, and not just continue to use those developed for thermal catalysis.

*e*-Chemistry thus requires a revolutionary rather than an evolutionary approach, in terms of capability to integrate, in a holistic perspective, many aspects from fundamental to applied chemistry and engineering for the deep transition required to move to a new FF-free sustainable future. Facing the transformation to a NZE society requires radical system shifts (concerted) in the same direction,[25,26] requiring thus novel assessment modes for their evaluation.[27] A concerted synergy between the development of novel routes and technologies and the parallel changes in the economic, industrial and societal systems is necessary, implying also to create the pathways by which this concerted mechanism can be achieved. Therefore, a framework assessment of the transformation is required rather than to assess single specific technologies.[28]

For example, an analysis of the status of the studies on the economics of $CO_2$ utilization[29] clearly reveals the limit of application of economic assessment models not properly consider the on-going deep transition and thus not based on a framework assessment. This is also one of the reasons why very contrasting opinions exists on the opportunity or not to develop new routes to electrify the chemical production, and how to properly rank the priorities.

For example, a recent paper by Ueckerdt et al.[30] discussing the pro and cons of *e*-fuels (where this term indicates the synthetic fuels produced from electricity and $CO_2$ via water electrolysis) argued that "*e*-fuels' versatility is counterbalanced by their fragile climate effectiveness, high costs and uncertain availability". We feel that this conclusion on the negative aspects of *e*-fuels, and the preferences about alternative solutions, derives from a series of assumptions. In particular that i) it is necessary to produce $H_2$ via electrolysis and separate and concentrate/purify $CO_2$, ii) *e*-fuels can be produced only by power-to-X technologies (i.e., producing $H_2$ by electrolysis and its use for $CO_2$ conversion by thermocatalytic process), and iii) renewable electricity should be that in excess with respect to other uses and with an intermittent production. We feel that these limitations will be overcome within the next decades, as discussed later in this manuscript. When a deep transition occurs as that on-going, evaluations are strongly depending on the scenario assumptions and the capability to consider the possible technological developments, despite the uncertainty associated to them. Most of the scenario analyses are limited from this capability, especially when system changes are occurring as those ongoing.

On the other hand, making estimations, including the economic aspects, without considering properly the technological developments was proven to lead to uncorrected indications. This is well demonstrated by failing predictions of the cost of electricity produced by PV and wind, that about two decades ago was estimated about 5-10 higher than the effective cost currently available. Analyzing literature studies on the economics of $CO_2$ utilization[29] to produce $CH_4$ and $CH_3OH$ (mainly by power-to-X technologies) we remarked the presence of a very spread range of calculated costs, much broader than the possible uncertainty in cost estimations. This indicates how to predict that a technology like the production of *e*-fuels will not have a role in the future scenario, based on only cost estimations, can be very dangerous. On the other hand, this evidences the difficulties in making an estimation on technologies still to be fully developed.

Although all predictions about the future scenarios are strongly dependent on many assumption, quite difficult to proven and with a high degree of uncertain, we believe that the conclusions made by Ueckerdt et al.[30] on negligible climate mitigation effectiveness of *e*-fuels depend on adopting a pessimistic scenario strongly affected from a cost analysis taken from literature and not considering in the right perspective the potential of technological development and innovation. In general, we could remark the need to i) account the deep transformation occurring and ii) identify the scientific and technological gaps to overcome in order to implement this transformation.

From the catalysis perspective, it is thus necessary to understand the directions and trends offering new opportunities, but also to identify properly the limits and gaps as well as the crucial issues to overcome.[31-33] Also in terms of catalysis technologies, it is necessary to address the limitations remarked above, for example the possibility to have photoelectrocatalytic (PEC) cells which are able at the same time to

1. convert directly $CO_2$ from diluted streams without need of concentration/purification (by integrating suitable membranes, for example,[34] or combining with integrated electrochemical retention of $CO_2$)[35],

2. operate directly with solar light 24h (by integrating in the PEC cell a redox storage for a temporal decoupling of the redox processes requiring light and those for the production of e-fuels)[36,37]

3. produce directly in a single cell the *e*-fuels/*e*-chemicals from $CO_2$, water and light.[32]

Research in these directions as well on the fundamentals aspects which differentiate reactive from thermal catalysis is still quite limited. By a proper intensification and focus of the R&D, we believe that within 10-15 years technologies overcoming the above limitations become feasible. However, it is essential that research will address the proper gaps and limits, with a better focus on crucial aspects to solve.[10] Also scenario analyses, when a deep transformation is involved, should be focused at identifying the future technologies and to assess the possible directions from this perspective, and whether the existing scientific and technological gaps can be bridged.

**Scope and Limits.** This perspective paper aims to contribute in the analysis of the prospects, limits and gaps from the viewpoint of the needs of scientific and technological developments in the field of catalytic technologies to realize an *e*-chemistry. This analysis is preceded by a short assess of the feasibility and timing to realize an *e*-chemistry as a necessary element of the transformation to meet the NZE targets by year 2050, when a proper R&D effort will be dedicated.

However, the aim is not to discuss in depth the different opinions in literature regarding the need to transform petrochemistry into *e*-chemistry.

Discussion on the needs to realize an *e*-chemistry will not specifically address the issue of scalability of the technologies discussed. In fact, the scope of this paper is oriented to a long-term vision of the future technologies needed to realize an *e*-chemistry (unexplored reaction pathways) rather that to analyze gaps in the currently available technologies. It is important to identify these future technologies to prepare all the scientific and technological bases for their realization, whether a deep discussion of technology readiness/development is related to current technologies.

The above discussion was mainly focused on electrocatalysis. Other catalytic technologies using RES, such as photo- and plasma-catalysis, or catalysis with microwave or other radiations such as magnetic heating will be also important to implement an *e*-chemistry. However, the latter two technologies (microwave or magnetic heating) change the mechanism of heating, but the process remains a thermal process, with all the related intrinsic thermodynamic limits. Photo- and plasma-catalysis instead share with electro-catalysis the presence of quite reactive species, as electrons, holes, radicals and for this reason lumped together as reactive catalysis, in contrast with the traditional thermal catalysis.[10] However, from an application perspective, electrocatalysis has a more mature stage of development, and advantages of higher productivities, efficiencies and process intensification.[8] In addition, it will take advantages of the increasing experience on fuel cells and electrolyzers, as well of commercial electrocatalytic processes (chloro-soda, adiponitrile, etc.). Finally, by using stacks it is possible to obtain high productivities per reactor volume. All these aspects will make the scale-up of the electrocatalytic processes faster. Electrocatalysis likely will be thus the first technology to be applied industrially for the development of new routes for *e*-chemistry. We have thus focused discussion here on electrocatalysis, but the relevance of other routes (in particular photo- and plasma-catalysis) need to be taken into account. However, it is also necessary to note that a proper comparative analysis of electro-, photo- and plasma-catalysis, and of pro/cons of these different technologies, is largely missing in literature. Moreover, data are reported in a way hard to be properly compared, for example in terms of productivity, energy efficiency, selectivity. Thus, an effort in this direction is necessary.

Note finally that we do not address here the role of bio-based chemicals in the development of a FFs-free chemical production, because has been already discussed elsewhere.[38-40] However, the bioroutes have often a still significant footprint and are not designed in most cases to integrate RES in the production.[41,42] Thus, a transition to a fossil-free *e*-chemistry for a NZE target would require to reconsider these routes in terms of integration with RES and with technologies, such as electrocatalysis (as discussed later), representing the link to effectively integrate RES in the process.

Biocatalysis instead will certainly play a significant role in the future *e*-chemistry, although the challenges of i) performance, ii) process costs and iii) process intensification must be solved to expand from dominant pharmaceutical and fine chemical applications to the production of bulk-chemicals.[38-40] Also in this case, the synergy and interface with electrocatalysis and other catalytic methods based on the use of RES have to be analysed to define the optimal paths for the future *e*-chemistry. However, this is a largely unexplored area.

## ■ FEASIBILITY AND TIMING FOR *e*-CHEMISTRY

There are many different opinions on whether an *e*-chemistry can be feasible, and when it could be effectively implemented on a large scale. Regarding timing, a question is when we would consider realized the implementation of the transition from petro- to *e*-chemistry.

The timing of a transition is when the new technologies (for *e*-chemistry) prevail over those traditional (currently in use) at the stage of planning new investments in industrial chemical production. In fact, every massive change of a production system requires time to be completed and there is an unavoidable period in which the new and old technologies will operate in parallel. The transition is realized when the novel technologies will start to be introduced for new plants. The further years are due to the time necessary to complete the switchover.

There are clear worldwide indications that investments on new plants and technologies based on FFs are significantly decreasing. Already large consulting companies such as McKinsey, as commented later, evidence the reduced attractiveness to invest in current petrochemistry processes based on FFs.[43] Deloitte, another major consulting company, also advised about the changing petrochemicals landscape and that petrochemicals industry is at a crossroads of major structural shifts.[44]

Despite the great uncertainty in predicting future, these signs and indications from consulting companies allow us to assume that for year 2050 the still likely use of FFs will be mainly a consequence of to this switchover period rather than of the profitable use of the current petrochemistry technologies (or slightly improved), especially in investing in new plants. Understanding this difference is crucial for a proper scenario analysis and prediction of the technological landscape in year 2050, particularly in geographical regions pushing the NZE transition, such as Europe. However, a prerequisite to realize this scenario is that the R&D investment should be increased, with the identification also of the implementation mechanisms able to focus the R&D activities on the key fundamental and technological aspects leading to an acceleration of the innovation.

An often-posed question concerns the availability of the needed green electricity for the transformation of petro- to *e*-chemistry and its cost. The McKinsey report "Pluggin in: What electrification can do for industry"[13] indicates that "renewables could produce more than half of the world's electricity by 2035, at lower prices than fossil-fuel generation". Many other reports are in line with this estimation. FFs are no longer considered as a privileged energy source in terms of costs, but their still dominating monopoly character determines a market controlled by other factors than those associated to a truly competitive economy, as shown in the second half of 2021.

Thus, to overcome the dominant use of FFs is a strategic direction and not only relevant to reduce greenhouse gas emissions.[45] Substituting the use of FFs is thus the result of different converging elements, from cost advantage, drastic cut in greenhouse gas (GG) emissions (and associated avoided costs) and improved energy geopolitics. It is the combination of these elements making irreversible the transition, and thus also petrochemistry should reinvent itself in this direction. McKinsey, a major consulting company, already some years ago indicated that petrochemistry has lost the window of opportunity as advantaged feedstock provided and therefore it needs to reinvent themselves.[43]

In petrochemistry, less than half of the FFs input (accounting for > 90% of input of the chemical sector) is used as feedstock (e.g., as carbon source), while the remaining part to produce the necessary energy for the chemical processes. Global fossil fuel consumption corresponds about 137,000 TWh and by considering that around 14% and 8% of total primary demand for oil and gas, respectively, is related to chemical production,[46] the latter uses globally account for about 12,000 TWh equivalent of FFs. Global amount of electricity generated from renewables (in 2021) was estimated in 8300 TWh,[47] but with the introduction of the new technologies for *e*-chemistry it is expected to increase the efficiency of energy use, thus reducing the energy consumption.[47] Clearly, actual renewable energy production is already in use for different applications, but by year 2050 the production of RES is expected to increase by a factor of 3-4 arriving up to 90% renewable share in electricity, with also a reduction in the total final energy consumption from 378 EJ (in 2018) to 348 EJ (in 2050).[48] In addition, the introduction of technologies such as that (indicated above) of PEC devices with integrated redox storage, to enable their continuous use overcoming intermittency of RES (one of the current main drawbacks), offers an additional path to produce *e*-fuels/*e*-chemicals using directly solar light.

In a transition path to 2050, FFs substitution in chemical production can be first realized by replacing their use as energy source (the so-called electrification of the chemical production) and then as carbon source, by introducing technologies for efficient closure of the carbon cycle coupled with a rational use of biobased resources.[49] The carbon recycle introduces energy efficiency, besides a better use of resources. The energy efficiency in recycling $CO_2$ to methanol, for example, is potentially up to over 80%,[50] while considering the different steps necessary to produce methanol from oil and the losses related to the extraction and transport of oil, the energy efficiency from fossils is on the average lower than 50%.[51,52] This is not the efficiency of the single process of methanol production (what often considered), but the global efficiency which accounts for the many steps necessary to arrive to methanol with the current route.

Converting $CO_2$ directly to acetic acid by an electrochemical process would further reduce the number of steps (a main industrial route involves the carbonylation of methanol) and improve the overall energy efficiency of the system.[53] The minimum process energy divided by the total process energy input is about 27% for acetic acid in the conventional route,[54] while the electrochemical route potentially can reach energy efficiencies above 60%, drastically lowering the carbon footprint. This concept is exemplified in Figure 1 reporting a Grassmann-type diagram of indicative comparison of exergy in the multistep conventional process to produce acetic acid using FF sources and the direct electrocatalytic route of $CO_2$ conversion to acetic acid with in-situ water electrolysis.[55] The difference between energy input (as sum of raw materials and fuel input) and the final work potential of the target product represents the sum of internal and external exergy losses, and those related to steam export, which can be hard to be utilized in distributed approaches. This resulting decrease of the carbon footprint is higher than 70%.

FIGURE 1 HERE

The great potential of these *e*-technologies is to introduce process intensification by reducing the number of steps, potentially lowering fixed and operative costs, and introducing new modalities of production with a better use of the local resources (distributed production). The full value chain of chemical production and the strong nexus with refinery would be changed in passing to an *e*-chemistry model. This is the perspective required for the analysis of the feasibility and impact.

We may conclude that in the frame of the proposed high tech scenario, the potential to substitute FFs in the chemical production with renewable energy and alternative C sources exists. This objective would correspond to a better use of the resources changing from a linear to a circular economy model. The impact is potentially to lower rather than increase the production costs as instead often claimed. Saygin and Gielen,[56] for example, estimated that "achieving full decarbonisation in this sector will increase energy and feedstock costs by more than 35%". Shreve[57] indicated that replacing fossil fuel power systems in the United States could cost up to $4.7 trillion. Markandya et al.[58] instead remarked that "the reduction of pollution and climate impact through rapidly increased use of renewable energy by 2030 could save up to USD 4.2 trillion per year worldwide." These are few among many other examples of the quite contrasting indications reported in open literature on the feasibility and costs of substituting FFs. In our high tech scenario, where an intensified R&D will allow to solve current technological limitations, the

innovation which brings from the transition to an *e*-chemistry will result in an overall reduction of the costs, in addition to an increased sustainability and reduced impact on the environment.[55]

In the past, when a proper stimulus of R&D was given, for example when the strong limitations on car exhaust emissions (Clean Air Act) were introduced around year 1970, the results were largely beyond expectations. Car exhaust treatment technology realized what was indicated as impossible to achieve. In addition, the current petrochemistry based on olefins was largely realized in few years around year 1960, due to a series of converging driving elements.[23] Also in this case, transformation occurred much faster than predicted. Therefore, history teaches how major technological changes, as those necessary to implement an *e*-chemistry, may occur despite many negative predictions. The scenarios for 2050 predict completely different situations in terms of reduction of GHG. We believe that it would be better to analyze what are the gaps and limits to implement a possible scenario, rather than to long discuss whether could be realized, using different assumptions all difficult to prove to be the most correct.

The key question thus remains how to develop the technologies which are needed, accelerating their discovery and industrial implementation. Catalysis, being a key element in most of these technologies, should follow the same trend and thus the question is how to accelerate the progress necessary in catalysis for *e*-chemistry, rather than to discuss whether this petro- to *e*-chemistry transition is feasible. We believe that many elements indicate that an irreversible transition towards the realization of an *e*-chemistry is already started.

## ■ LIMITS AND GAPS IN CATALYSIS FOR *e*-CHEMISTRY

The production modes in petrochemistry are rigidly hierarchical with few building blocks (mainly light olefins, aromatics and syngas/methanol) requiring a sequence of steps, often many, to obtain the final chemicals for industrial (polymers, synthetic fibres and rubbers, solvents, etc.) or consumer uses (detergents, drugs, fertilizers and pesticides, paints, etc.). This production scheme is largely associated to the concept of scale economy, e.g., the need to develop large-scale centralized plants and thermal processes.[23] This model of chemical production has many limits, from the significant local impact on the environment to the intrinsic low flexibility and adaptability instead required due to an uncertain future. Chemical plants are designed for a utilization factor typically above 90% to operate economically. The global ethylene production plants decreased to an average 82-83% in the last years and it is predicted to remain lower than 90% in the next decade.[60] In these conditions, economic margins for the production are very low, or even negative. This is a general situation for petrochemical production (ended the windows of opportunity)[43] and it will be accentuated in the future, needing to change the production model from centralized (few very large sites) to a distributed model, more flexible and strongly reducing costs and impact of transport/distributions.[60] In addition, a distributed model offers the integration with the local resources rather than along the value chain (creating also new opportunities for symbiosis and investments). All these elements create a competitive environment based on innovation. A distributed production requires efficient small-medium scale plants well integrated with the territory and the local resources/needs, with a modular plant scheme allowing faster time to market and great flexibility of operations. *e*-Chemistry technologies should have these characteristics.

Therefore, the remark often made that it is not possible to produce large-scale building blocks as ethylene (typical size of steam crackers to produce ethylene goes from 200 to over 1000 ktons/y of ethylene) by electrocatalysis (or other routes based on RESs) is mispresented. In *e*-chemistry the model of production is changed with the target of small-medium scale productions tailored for the local production needs. It avoids distribution/transport on a large-scale. The aim is the direct conversion to the final product (or at least to strongly reduce the number of steps), avoiding fragmentation of the production in a long sequence of steps. Ethylene is a raw material for polymers (polyethylene), but also the building block for a large series of other chemicals (used mainly for other polymers) such as vinyl chloride, ethylene oxide, vinyl acetate, etc. Thus, technologies for the direct ethylene production should be used for polyethylene manufacture, while other chemicals derived from ethylene should be ideally produced directly rather than via ethylene as intermediate.

Many well-established large-scale chemicals are produced with a sequence of steps which can be potentially drastically reduced. As an example, phenol production is realized commercially with the conversion of benzene to cumene, which is converted to cumene hydroperoxide then decomposed to phenol and acetone. However, the yield is low, about 8% (on the whole process), due to the critical step of the intermediate cumene hydroperoxide production. The direct one-step synthesis of phenol from benzene and $H_2O_2$ is potentially competitive but suffering of hydrogen peroxide cost.[61] $H_2O_2$ could be produced electrochemically and thus a novel electrochemical route to directly produce phenol from benzene is potentially feasible. It may be a hybrid system with production of $H_2O_2$ electrocatalytic combined with in-situ catalytic conversion, or better by direct electrocatalytic benzene hydroxylation using hydroxyl species generated at the anode. There are many open questions, from the type of electrolyte and electrode to use to the control of multiple hydroxylation (phenol is more reactive than benzene towards the insertion of a further hydroxyl groups, but there are strategies to control this issue by inhibiting the further reactivity of phenol).[61] However, studies in the field are extremely limited. The production of $H_2O_2$ by electrocatalytic routes is a topic of growing interest,[62,63] but not the direct electrocatalytic hydroxylation of benzene, or the hydroxylation of other substrates. However, papers are available on the catalytic hydroxylation of benzene with $H_2O_2$.[64-67] Benzene direct electrocatalytic hydroxylation is one of the routes to explore for an *e*-chemistry, but which has been not yet investigated. Benzene current source derives from FFs, mainly as a product of the refinery reforming process. However, in a future *e*-chemistry it could be produced from lignin.[68]

The example above introduces the concept of extending the use of electrocatalysis by addressing complex syntheses of chemicals by combining electrocatalysis of small molecules ($CO_2$, $H_2O$, $N_2$, $CH_4$, the latter from biogas sources) and of biobased molecules. The integration of these two separate worlds is the grand challenge for *e*-chemistry and *e*-refinery.[69]

Most of the electrocatalysis studies instead address simple reactions as the two-electron reduction of $CO_2$ to CO or to HCOOH.[70-73] Also in $CO_2$ electrocatalytic conversion, the challenge is instead to realize complex conversions of $CO_2$ leading to multi-carbon (C2+) products and to the intermediates needed to build a petrochemistry-equivalent framework, as commented above. Target is to produce directly both the base chemicals which can be used then as raw materials for the current chemical production (light olefins, for example) or even better produce directly (in one-step) more complex molecules.

The direct synthesis of multi-carbon products is a topic of growing interest in $CO_2$ electroreduction,[33,74-76] but mainly from an academic perspective rather than as part of a strategy to build the new *e*-chemistry. For example, being intrinsically simpler the electroreduction of $CO_2$ to CO than the formation of C2+ products, a large part of the studies focused to the $CO_2$ electroreduction to CO, with the justification that the performances (Faradaic yield, productivity) are better, and CO could be then used in combination with $H_2$ to make a variety of other chemicals, via methanol or Fischer-Tropsch catalytic processes. In general, the productivity of $CO_2RR$ to C2+ products (especially C2+ hydrocarbons) is still low, although significant progresses have been made recently and now Faradaic selectivity and productivity/current density (to ethylene and ethanol, in particular) have reach levels in some cases which allow to consider a possible industrialization.[74-76] However, still a gap exists between production of syngas by co-electrolysis on solid oxide electrolyzer cell (SOEC). The latter is a better solution with respect to the separate production of CO from $CO_2$ and $H_2$ from $H_2O$ in PEM-type electrolyzers, although SOEC still shows relevant issues in terms of cost/performances, reliability and durability.[77] After producing the syngas, with eventual adjustment of the $CO/H_2$ ratio, a compression and heating could be necessary, then one or more thermocatalytic steps of conversion could be necessary. Two routes could be possible to obtain olefin: via Fischer-Tropsch (FT) process (in the modified version FT to olefins, although performances are still unsatisfactory notwithstanding the recent progresses)[79] or via intermediate methanol synthesis followed by methanol to olefin conversion (followed by further C4+ cracking unit).[80] In both routes a broad distribution of products is obtained. If the overall efficiency, especially energetic, is accounted for these multistep processes, and the needs of complex downstream separation units which also makes costly the development of small-scale distributed applications, the direct electrocatalytic production of olefins appears as a preferable route.[81] However, the multistep power-to-olefin route (syngas by co-electrolysis, then thermocatalytic steps) is more mature in terms of implementation. In this case, a decrease of the carbon footprint could be realized through electrification of the thermocatalytic process, e.g. using electricity to provide the heat of reaction.[82]

A wider range of possibilities exists (than forming C2+ production from $CO_2$) which can be used to build new routes for *e*-chemistry. An example is the electrocatalytic reductive coupling of $CO_2$ to oxalic acid,[83,84] which can be further electro-reduced to a range of valuable chemicals for polymerization like glycolic acid, creating a new C2 value chain from $CO_2$.[85-87] Further reduction of glycolic acid may lead to produce ethylene glycol (a main intermediate for polyesters), thus a new path with respect to the current one via ethylene and ethylene oxide. This electrocatalytic chemistry is investigated in the EU project OCEAN (Oxalic acid from $CO_2$ using electrochemistry at demonstration scale, grant 767798). Oxalic acid can be electrocatalytic reduced to glyoxylic acid and further to tartaric acid,[88] opening a new path to C4 chemistry from $CO_2$.

The electrocatalytic production of acetate/acetic acid from $CO_2$[53,89] is also opening new possibilities, not only for the use of acetic acid itself (as solvent) but also to produce electrocatalytically a range of interesting products. One of them is the combination of the electrocatalytic production of acetate from $CO_2$ and its reaction with (bio)ethanol to form ethylacetate, a green solvent of large use. The electrocatalytic reactor is used to produce ethylacetate both at the anode side by anodic oxidation of ethanol, and at the cathode side, through in-situ catalytic reduction of $CO_2$ to acetate which reacts with ethanol to form ethylacetate. Therefore, the same product is produced at both electrode sides. This is an interesting example of coupling $CO_2$ electroreduction chemistry to the use of chemicals from biorefinery. This electrocatalytic chemistry is investigated in the frame of EU project DECADE (Distributed chemicals and fuels production from $CO_2$ in photoelectrocatalytic devices, grant 862030). Another possibility is to explore a similar chemistry but finalized to the direct synthesis of vinyl acetate. No studies in this direction are available, but in an old patent[90] an electrochemical process to produce vinyl acetate from ethylene and anolyte acetic acid is claimed.

There is a much wider range of possibilities, for example to develop paired or tandem electrocatalytic conversions.[91-94] One example is the oxidation of glucose to gluconic acid on the anode side and the hydrodeoxygenation of gluconic acid to adipic acid, allowing the one step electrosynthesis of adipic acid, a large-scale monomer, from a biobased platform molecule (glucose). This is one of the target reactions of the EU project PERFORM (PowerPlatform: establishment of platform infrastructures for highly selective electrochemical conversions, grant 820723) aiming to develop a small-scale pilot reactor. TERRA EU project (Tandem electrocatalytic reactor for energy/resource efficiency and process intensification, grant 677471) instead investigated the tandem electrosynthesis from C6 and C5 sugars (respectively) of the two monomers for PEF (PolyEthylene Furanoate): 2,5-Furandicarboxylic acid and ethylene glycol.

These are some examples of the on-going studies, involving various industries, to build new routes based on electrocatalysis for an *e*-chemistry, in some case also at pilot/demo scale. Note that even if the possibility to make a more complex chemistry in the electroreduction of $CO_2$ is known from nearly 15 years[95] only in the recent years started a more systematic study in this direction. Similarly, while organic electrosynthesis is a well-established scientific area,[96-99] and electrochemistry (of fuel cells) is known from a century,[100] the realization of an *e*-chemistry poses new problems and challenges. Electrosynthesis is one of the organic synthesis tools, designed mainly for specialty and fine chemicals, with different typologies of reactions (C-H activation. cyclization, dehalogenation, carboxylation, coupling, etc.). However, it is typically not suited for bulk chemical syntheses.

Now many new possibilities started to be opened by recent studies in electrocatalysis. The electrosynthesis of ethylene and propylene oxides is an interesting new area of investigation,[101] although already two decades ago the electrosynthesis of alkene oxides (and glycols) from alkenes has been reported.[102] Sargent and coworkers[101] epoxidized electrochemically ethylene and propylene to the corresponding epoxides (EO and PO, respectively) at industrially relevant current densities with Faradaic (electron-specific) selectivities ~70%. They coupled an electrochemical flow cell to homogeneous reactions for an overall stoichiometry (for ethylene oxide)

$$C_2H_2 + H_2O \rightarrow C_2H_2O + H_2 \qquad (1)$$

although chlorine is used to mediate the process. At the anode, the dissolved olefin is converted to the chlorohydrin, and the hydroxide is produced at the cathode along with $H_2$. Then, in a downstream process the anodic stream containing chlorohydrin [$RCH(OH)CH_2Cl$, where R can be H for ethylene or $CH_3$ for the propylene case] and HCl are mixed to the cathodic stream (containing $OH^-$) to generate the epoxide, $Cl^-$ and water. The authors also propose to couple the system to the electrocatalytic conversion of $CO_2$ to ethylene, realizing an integrated $CO_2$-to–ethylene oxide process. Sargent and coworkers[101] suggest that this process could be scaled to produce ethylene oxide (EO) at a comparable cost with current industrial practices. Although some open questions exist regarding energy intensity, safety of operations ($Cl_2$ is a very toxic gas, although chloro-alkali technology is one of the oldest electrochemical processes) and stability of materials,[19] the above results open new possibilities.

Another example of unconventional application of electrocatalytic routes is in the selective hydrogenation of acetylene.[103] Acetylene hydrogenation is an important industrial reaction for purification of ethylene streams. Here the authors[103] proposed this technology as a final step in a non-oil route to produce ethylene from natural gas via methane pyrolysis. They reported a high Faradaic efficiency of 83.2% for ethylene with a current density of $29\,\text{mA·cm}^{-2}$ on Cu electrocatalysts, due to strong acetylene chemisorption which limits the side hydrogen formation. This concept of competitive chemisorption to enhance selectivity in electrocatalytic reactions has a more general validity in electrocatalysis.

Electrocarboxylation of olefins and diolefins[104] (ethylene and butadiene, especially, both which can be produced from bioethanol)[105] or of aromatics is another route of interest to build an *e*-chemistry.[106,107] A variety of valuable intermediates could be produced by this method.

When other reactants are present, for example nitrate, the spectrum of the possible products further enlarges. One example is the electrochemical synthesis of glycine from oxalic acid and nitrate,[108] rather than the electroreduction of oxalic acid to glycolic and glyoxylic acids discussed before. Glycine is one of the base aminoacids.

Most of research in electrocatalysis instead focus on few limited reactions ($CO_2$ conversion, $N_2$ fixation, water electrolysis or $H_2O_2$ production, conversion of few biobased platform molecules), and in particular on the electrocatalysts development, including their design criteria, instead to explore the large range of possibility to build an *e*-chemistry and the interface with biotechnological paths (coupling biorefinery and *e*-chemistry).[109]

Therefore, there is an increasing interest to build new possibilities and routes for *e*-chemistry, especially at the interface with biobased platform molecules, but, on the other hand, research is still rather scattered and a very systematic exploration of all the possibilities is missing. It is necessary to attempt to build a new chemical production framework based on the electrocatalytic framework which combines the conversion of small molecules ($CO_2$, $N_2$, $H_2O$, $CH_4$, the latter derived from biogas) with those obtained from the electrocatalytic conversion of biobased platform molecules. The reactions to explore regard both i) the coupling of in situ generated products with those of biobased molecules (as the cited reaction of in-situ generated acetate to form vinyl acetate o ethyl acetate), and ii) the large-range of novel possibilities by using the in situ generated intermediates in the electro-transformation of the indicated small molecules. Figure 2 summarizes the concept that by combining primary electrocatalytic reactions of small molecules (by using final products and active intermediates, in a combined approach), to the electrocatalytic conversion of bio-derived products, it is possible to produce a large range of products forming the skeleton of a new chemical production.[69,109] Reactions to explore include i) direct coupling of in-situ generated intermediates, ii) tandem reactions using in-situ generated chemicals, and iii) the eventual coupling between electro- and heterogeneous catalysis.

By properly combining the primary reactions of electro-conversion, a larger range of products can be derived, potentially meeting the needs for a novel *e*-chemistry in combination with the other paths described before. This approach extends the concept of tandem electrocatalytic processes discussed before.

In tandem or paired reactions, the concept is the valorisation of both cathodic and anodic reactions to form added-value products. In water splitting, but similarly in most studies on $CO_2$ or $N_2$ electroreduction ($CO_2$RR and NRR, respectively), $O_2$ is produced at the anode site: the oxygen is most of the practical cases cannot be used, and is released to atmosphere, and even when used, it is a low-value chemical. In addition, oxygen evolution reaction (OER) is a four-electron reaction and it is typically the slow step of the electrocatalytic process, which determines most of the energy losses due to overpotential. Moreover, to produce $H_2O_2$ by water oxidation instead $O_2$ (a two rather than a four electron reaction) is an interesting target.[62,63] $H_2O_2$ is a valuable oxidant in many selective oxidation processes (as discussed later) and in environmental applications. There is thus a rising interest on this process which represents not only an alternative to OER but also a synthesis process for $H_2O_2$ as industrial selective oxidant, even if attempts to produce $H_2O_2$ electrocatalytically are known from decades.[110] $H_2O_2$ is already production in Mtons scale by anthraquinone route as selective oxidant in commercial processes (propylene oxide, caprolactam, catechol and hydroquinone syntheses), for soil and water remediation uses and as a versatile bleaching agent.[23]

$H_2O_2$ formation is half the reaction of the most studied oxygen evolution reaction (OER), the anodic reaction in water electrolysis systems.[111] A large gap between current performances and those necessary for industrial exploitability, however, is present. Also, for this electrocatalytic synthesis of $H_2O_2$ there is the need to develop more solid bases for the electrocatalysts design. For example, by theoretical modelling the binding strength of the reaction intermediate *OOH to the catalyst surface is indicated as the key parameter for controlling the catalyst performance. However, recent studies, for example on single-atom electrocatalysts, are in contrast with these mechanistic indications.[112] Recent advances in the understanding and design of the gas–liquid–solid three-phase architecture have led to significant progresses in obtaining a high selectivity (83–99% current efficiency) combined with high current density and stability.[113] By using a superhydrophobic natural air diffusion electrode (NADE) to improve the oxygen diffusion coefficient at the cathode as compared to the normal gas diffusion electrode (GDE) system, Zhang et al.[114] showed that it is possible to largely increase $H_2O_2$ production rate and oxygen utilization efficiency. These examples show how the deep understanding of the engineering aspects at the electrode nanoscale is the crucial factor in determining the behaviour, and not only the nature, of the electrode itself.[115] Quite similar aspects are also strongly determining the behaviour in NRR.[116] Produce $H_2O_2$ rather than $O_2$ in water splitting or other electrocatalytic reduction reactions ($CO_2$RR and NRR) has thus the advantages of making a 2e⁻ rather than a 4e⁻ process (with advantages in terms of rate and reduced overpotential), and obtain a higher value product. $Pd^{\delta+}$ clusters ($Pd_3^{\delta+}$ and $Pd_4^{\delta+}$) onto mildly oxidized carbon nanotubes (containing controlled defects) were recently shown to allow nearly 100% selectivity in $H_2O_2$ formation with low overpotential and high mass activity.[117] Therefore, significant progresses have been made recently on the electrocatalytic production of $H_2O_2$.[118-122] However, even if $H_2O_2$ is used in many chemical applications, the volume of the potential commercial market is largely lower with respect to that it would be needed for potentially very large-scale reactions as $H_2$ production by water electrolysis, $CO_2$RR and NRR. In addition, in many applications of $H_2O_2$, for example in its industrial use in selective oxidation reactions, the concentration and solvent requirements do not fit well with those produced in the electrocatalytic $H_2O_2$ synthesis.[23] To find an alternative optimal reaction to OER remains thus still a challenge. Some further aspects will be discussed later. Note also that the increasing trend to operate in non-protic solvents to limit side reaction of $H_2$ formation in $CO_2RR$ and NRR further stress the need to identify valuable anodic substitutes to OER and $H_2O_2$ production as well.

Coupling between electro- and heterogeneous catalysis is another area of interest to remark. The development of hybrid catalysts/reactors combining electro- and heterogeneous (or eventually homogeneous) catalysis is an area still largely unexplored which represents a specific opportunity. On the contrary, hybrid electrocatalytic/biocatalytic systems have been more systematically explored.[123-125]

Figure 3 reports a framework scheme of the possibilities derived by developing hybrid electro- and biosynthetic catalytic pathways in $CO_2$ conversion. The range of possibilities offered from this symbiosis is evident, further enhanced by considering that enzymes for converting other small molecules ($N_2$, $H_2O$ and $CH_4$) are also known. A critical challenge would be the transfer of energy and reactive compounds from the electrocatalyst to the cascade (or integrated) biocatalytic process. Three types of configurations can be possible: i) direct attachment or ii) indirect attachment of microorganisms on the electrode, and iii) sequential (physically separated, cascade) reactions at the electrode and the microorganisms. The same configurations can be present also in integrated electro- and heterogeneous/homogeneous catalysis. Each type of configuration has pro and cons.

FIGURE 3 HERE

In addition, direct or mediator-assisted electrocatalysis is possible. The use of redox mediators (typically organic compounds, but not limited to them) is common in organic electrosynthesis.[126] In the indirect electrocatalysis the electron transfer step is shifted from a heterogeneous process occurring at an electrode to a homogeneous process using an electrochemically generated reagent ("mediator"). The latter is involved in a reversible redox couple initiated at the electrode. The use of these mediators can better control the selectivity, the stability of the electrodes and can overcome kinetic inhibitions. Several new redox mediators have been developed recently, offering a range of possibilities including unconventional types of operation such as double mediatory systems for biphasic media and enantioselective mediators.[126] Redox mediators play a crucial role in electroenzymatic reactions (co-factor regeneration). Indirect electrochemical conversions are hybrids between direct electrochemical conversions and homogeneous redox reactions. The electrochemical generation and regeneration of this activated species can either proceed in situ in the same electrochemical cell (in-cell process) or in a separate cell (ex-cell process), allowing also to realize a spatiotemporal decoupling of the processes. Tetramethylpiperidine N-Oxyl (TEMPO), Phthalimide N-Oxyl (PINO) and related N-Oxyl species are widely used as mediators in electrocatalytic reactions.[127] N-oxyl compounds undergo facile redox reactions at electrode surfaces, enabling them to mediate a wide range of electrosynthetic reactions. Oxidation of aminoxyls, such as TEMPO, generates oxoammonium species that commonly promote hydride transfer from organic molecules, such as alcohols and amines, to obtain hydroxylamine. While often used in organic electrosynthesis, they are still scarcely explored in larger-scale processes, for issues of separation and stability that often become the discriminating elements for industrial processes.[128] On the other hand, by using heterogenized mediators, and by designing the electrocatalytic reactor with integrated membranes it is possible to overcome, in principle, these issues and exploit the benefits of mediated electrochemical reactions.

## ■ KEY DIRECTIONS TO BUILD AN *e*-CHEMISTRY

Above discussion evidenced a series of key directions to build the new catalysis required by *e*-chemistry. It is useful to summarize them to remark some of the areas on which R&D should

be pushed:

1. The development of PEC devices integrated with redox mediators that allow to make the direct solar-to-fuels/chemicals conversion with continuous (24h) operations, because the redox mediators store the energy when light is present on one electrode side, allowing to operate on the other electrode side in electrocatalytic continuous modes. The spatiotemporal decoupling of the processes in PEC devices overcomes the limits of intermittency (with associated costs and issues) offering also a new possibility to design the devices and to overcome other limits such as insufficient current density for industrial operations.

2. The study of the benzene direct electrocatalytic hydroxylation to phenol, which can be also considered a model for a wider range of novel *e*-chemistry routes.

3. Exploring more systematically the novel electrocatalytic routes by coupling the products of the conversion of small molecules ($CO_2$, $N_2$, $H_2O$) with the in-situ functionalization of biobased molecules (or using the products formed in integrated hetero-, homo- or bio-catalysis cascade processes). The examples discussed regard the reaction of acetate (formed from $CO_2$) with bioethanol or bioethylene to form ethylacetate (green solvent) or vinylacetate (monomer). This is a new area to explore opening many new possibilities, from the production of glycine (aminoacid) to the large range of novel paths by amination, nitration, selective epoxidation or oxidation, etc. The possibility to revise the old chlorohydrin synthesis to produce EO and PO was discussed as an example for novel electrosynthesis paths.

4. Investigating the electrocarboxylation of olefins and diolefins, or aromatics, as another route of interest to build an *e*-chemistry. A variety of valuable intermediates could be produced by this approach.

5. Starting to analyse the even more challenging, but conceptually possible, use of the reactive

intermediates formed in the electrocatalytic conversion of the cited small molecules to make new synthetic pathways for *e*-chemistry. The example of benzene hydroxylation falls within this area. Direct electrocatalytic amination (formation of C-N bonds) is an interesting possibility which can exploit also nitrogen sources as nitrate, also produced, for example, by plasma-catalysis from air.[129]

6. Exploiting the new possibilities offered from the electrocatalytic conversion of biomethane, which has been at large unexplored up to now.[130,131]

7. Exploring more systematically the many possible new paths, offering also process intensification, given by tandem and/or paired electrocatalytic reactions.[132,133] We have cited the examples of direct (one-step) glucose conversion to adipic acid, or the production of the monomers for PEF polymer synthesis at both sides of the electrocatalytic cell, a novel greener polymer substituting PET (polyethylene terephthalate), one of the largest polymer in current use.

8. Intensifying research on the direct synthesis of ethylene or methanol/ethanol by electrocatalytic reduction of $CO_2$, which were not discussed here being already discussed in many reviews, for example by Zhao et al.[2] and Ra et al.[16] Ethylene or methanol/ethanol can be then used as such or as building block for further *e*-chemistry. A large part of the research in these areas is focused on developing electrocatalysts with improved performances (especially Faradaic efficiency), and to identify the mechanistic features responsible for this improvement. Even if these aspects are relevant, those determining the industrial exploitability are often different from productivity and stability, such as the design of reliable and cost-effective electrodes and electrocatalytic reactors, with continuous operations, at high current densities. Often these aspects also determine the choice of electrode and operation conditions. The development of the electrocatalysts should be made under relevant conditions, while this is not the case of a large part of the current investigations. A better balance between the first type of academic studies and those more relevant for exploitability and integration within a general *e*-chemistry framework would be necessary.

9. Extending the investigation on the possibilities offered by C2+ (multicarbon) chemistry in the $CO_2$ electroreduction. We have discussed the possibilities by synthesis of oxalic acid, and its electro-conversion to form glycolic acid (for new polyesters) or other valuable chemicals, including ethylene glycol, a large-volume monomer. Oxalic acid can be converted to tartaric acid (via glyoxylic acid), giving rise to a new C4 path. Many other valuable paths are also part of the portfolio of electro-conversion possibilities of $CO_2$ to C2+.

10. Investigating with a broader approach the direct synthesis of ammonia (or derivatives, such as the synthesis of amination products) from $N_2$, another emerging electrocatalytic path to both fertilizers (perhaps the largest volume chemical) or N-containing chemicals.[134,135] Here it was also earlier evidenced[10,31] that the large research effort is improving too slowly from a more practical perspective, in part due to the use of approaches which are not accounting the difference between electro- and thermal catalysis. Methodologies (including theoretical) which are not able to catch the intrinsic difference in electrocatalysis are typically used. The result is to "demonstrate" a very large range of (contradictory) mechanistic features for relatively analogous electrocatalytic results. The use of different approaches, including biomimetic based on multi-electron/proton simultaneous transfer (requiring a different nature of the active sites) is a direction offering new clues to design electrocatalysts from a different perspective.[136]

11. Intensifying the investigation on the electrocatalytic production of $H_2O_2$.[110] Besides to the commercial interest of hydrogen peroxide as commented before, it represents half of the reaction of the most studied OER, the anodic reaction in water electrolysis systems. It is thus a two-electron path alternative to four-electron path to produce $O_2$ at the anodic part of the cell. It is a lower overpotential reaction, requiring less electrons and giving an added-value product, contrasted, however, from the easier decomposition. A general discrepancy was observed between the simple (H-type) electrochemical cells used in most of the studies, and those required for industrial practice. This aspect remarks that the use of appropriate electrocatalytic cell is crucial to obtain reliable indications. A large gap between current performances and those necessary for industrial exploitability is present, although this aspect is common to most of the current studies in electrocatalysis. The studies can lead to the understanding how to generate and use for different syntheses the electrogenerated oxygen active species.

12. Investigating more systematically how to develop electro-synthetic paths for bulk chemicals starting from biobased molecules. Studies in the electrocatalytic conversion of biomass-derived platform chemicals are i) still limited, ii) focused on few types of reactions (furfurals conversion, glycerol)[137] and especially iii) still unable to identify the rational bases to design the electrocatalysts. Although understanding on electrocatalysts is improving,[138] with the identification of various factors such as geometric structure of the active sites, presence of surface defects, size and shape of the nanoparticles, solvent, and electrolyte effects determining the behaviour, often the indications are specific for a defined catalytic system and results are in contradictions between different materials. Thus, remains difficult to define general aspects which allow to identify a priori the class of electrocatalysts to be used by hypothesizing the conceptual reaction mechanism. For complex electrocatalytic transformations (multi-electron/protons, when biomass-derived chemicals are involved, etc.) still exploration of new reactions remains largely phenomenological. General guidelines for catalyst selection which can restrict the field of investigation to a limited range of materials, as often occurs in the case of thermocatalytic reactions, are also not available. Theoretical approaches for catalyst selection are strongly depending on having established a precise mechanism of reaction, while the more effective electrocatalytic approaches have still to be clearly identified. We believe that even if progresses have been made, design in silico novel electrocatalysts is still not currently possible. A primary research objective should be thus the development of the fundamental framework knowledge to realize this objective. Reviews in the field, attempting to rationalize the research effort in electrocatalysis,[139] reported a status of the research allowing a better identification of the electrocatalysts, and more precise identification of the reaction conditions. However, a more intense effort, perhaps approaching from a different perspective, would be necessary to address properly the a priori design of efficient electrocatalysts.

Several electrocatalytic classes of reactions were identified by Tang et al.[69] to develop the bases for the *e*-chemistry outlined in Figure 2: i) C-N coupling, to form a range of N-containing chemicals, such as urea, amides, amines, and nitrile of widespread use as bulk, intermediates or specialty chemicals, for applications ranging from chemical production to fertilizers, agrochemicals, and pharmaceuticals, ii) selective hydrogenation or hydrogenolysis, with in-situ generation of $H_2$ or hydrogen equivalents ($H^+/e^-$), and iii) selective oxidation or oxidative dehydrogenation, also without addition of external oxidation agents, including $O_2$. However, the challenge is still to identify enough selective electrocatalysts to make industrially feasible the process. Tang et al.[69] provided a series of examples of electrocatalytic reactions for the three classes of above cited reactions. However, these examples are still far from industrial exploitability, and even within a homogeneous class of reactions, it is difficult to identify a rationality in the electrocatalysts design. Cardoso et al.[140] also discussed a series of redox and acido/base electrocatalytic syntheses, and a series of industrial examples of pilot development, although typically for specific cases and small productions. Moving from these examples related to organic electrosynthesis to the production of bulk chemicals by electrocatalysis is a challenge, which often requires to turn the type of approach used up to now.

Note that in addition to direct electrocatalytic syntheses, also indirect syntheses via redox mediators (TEMPO being one of the most used) are possible. Thus, various examples of organic electrosynthesis are known, but related to electrochemistry rather than to electrocatalysis. Addressing the challenge given from developing a full framework for *e*-chemistry (Figure 2) requires making a next step in combining catalysis to electrochemistry.

## ■ FUNDAMENTAL AREAS FOR AN *e*-CHEMISTRY

Research attention should be putted on specific fundamental areas, still not enough explored, in order to create the bases for *e*-chemistry. These areas should complement the development of new electrocatalysts and mechanistic studies, on which research attention is mainly focused currently. Among the different aspects to develop, the following can be highlighted:

- The design of advanced electrocatalytic reactors, which strongly determine the performances and behaviour, while still most of the studies are based on too simple reactors (like H-cells) not allowing to proper translate the results to realistic reactors closer to industrial exploitability.

- The study of multiphasic electrocatalytic reactors, an area that as outlined before is crucial to explore novel electrocatalytic paths, but scarcely investigated.[141]

- Understanding the difference, including in mechanistic aspects, between electrocatalysis (in general reactive catalysis) and thermal catalysis,[10] as the basis to develop new approaches and explore novel (complex) pathways. Current studies consider mainly the application of methods used for thermal catalysis to reactive catalysis, with still limited attempts to understand the difference in operations and type of active centers. Dynamics of polarized interface is crucial, also in determining selectivity[33] and significant reconstruction of active nanoparticles may occur upon application of a potential during electrocatalytic operations.[142]

- Electrocatalytic behaviour is determined by an intricate interplay between surface structure (both on the nano- and on the mesoscale), electrolyte effects (pH, buffer strength, ion effects) and mass transport conditions, a complex interplay still far from being completely understood.[7] A new electrocatalytic design, for electrolyte-less operations, allow to improve the behaviour by exploiting this interplay, but still systematic knowledge are limited.[143,144] A better understanding and a rational design of three-phase boundary in electrocatalysis can lead to an enhanced control of the performances.[144-147]

- Addressing the selectivity challenges of electrocatalysts. In several reactions, like $CO_2$ electroreduction, many reactions are possible at the applied potential, which includes the necessary overpotential to drive the reaction at sufficiently high rate. Developing an effective theory of selectivity in electrocatalysis is still at the infancy, even with the progresses in the field.[148-150]

■ **CONCLUSIONS**

The examples presented show that the potential and feasibility to develop a new *e*-chemistry exists, but it is necessary to push the research to address more specifically this challenge, exploring the various possibilities indicated and integrating in a holistic view the different developments to accelerate the discovery of new paths and the identification of the best options among the different possibilities. The new *e*-chemistry requires to turn the approach in the development of new paths, with a push towards new direct reactions which limit as much as possible the need of multistep processes realizing direct processes in small and efficient devices based on the use of RES. This is a grand challenge, which needs to be addressed accelerating the evolution in this direction. Although attention was given here mainly to electrocatalysis to focus the discussion, the importance of exploring also other routes of direct use of RES, such as photo- and plasma-catalysis, must be remarked again.

A general indication given is that to properly address the challenge of *e*-chemistry, there is the need to target complex (direct) syntheses by electrocatalysis. The question which may be posed is that what presented above is a dream (electro)catalysis and chemistry and it should be better to focus on simpler electrocatalytic reactions (two electrons, such as $CO_2$ to CO), because more complex multielectron/proton transformations are too challenging. The reply to this question is that it was scarcely attempted to make a rational approach to catalysis complexity, identifying the ways to proceed faster in targeting complex (direct) syntheses. Not addressing this question will delay the acceleration in the progress in this area which is crucial to meet the challenge of converting petro- to *e*-chemistry.

In conclusion, we believe that an *e*-chemistry (a chemistry using RES, closure of the carbon cycle and biomass as key elements to defossilize the industrial chemical production) is feasible (assuming a high-tech scenario), and this transformation from current petrochemistry is already started and will be largely irreversible. The nexus between refinery and chemistry is already changing, as a sign of this irreversible transformation.[151] However, acceleration of this process requires to turn the approach, with a broader vision of the future. For catalysis this will require to revise some of the fundamental bases and recognize that the new methodologies essential for *e*-chemistry need to approach catalysis from a wider and different perspectives, although based on the extensive background on knowledges on catalysis developed in the last decades.[152,153] In deep transitions, a synchronism between R&D and economics/societal changes is necessary. Thus, the development of novel sustainable process systems requires thinking at a multi-scale level identifying the energy-efficient and highly integrated systems deployed within local and regional contexts.[154]

While the transformation of the chemical industry to an *e*-chemistry is often considered to be motivated (only) by reducing the carbon-footprint, but at the "expense of higher production costs and unintended environmental burden shifting",[154] we would further remark that instead the push to innovation necessary will results in lower global costs, with a different model of chemical production, interaction with territory and society.[155]

For catalysis, *e*-chemistry is a grand challenge, offering the possibility to rebuild its role as key technology and develop new bases for understanding. This will occur only when the significant changes in the approach to catalysis required to face this transformation are understood and introduced in the scientific practice.

AUTHOR INFORMATION

**Corresponding Authors**

**Gabriele Centi and Siglinda Perathoner** − Departments ChiBioFarAm, ERIC aisbl, and CASPE/INSTM, University of Messina, 98166 Messina, Italy; orcid.org/0000-0001-5626-9840

**Other Authors**

**Georgia Papanikolaou, and Paola Lanzafame** − Departments ChiBioFarAm, ERIC aisbl, and CASPE/INSTM, University of Messina, 98166 Messina, Italy; orcid.org/0000-0001-8814-1972

**Author Contributions**

The manuscript was written through contributions of all authors which all equally contributed All authors have given approval to the final version of the manuscript.

**Funding Sources**

This work was made in the frame of the ERC Synergy SCOPE (project 810182), EU DECADE project (nr. 862030) and PRIN 2017 project MULTI-e nr. 20179337R7 and $CO_2$ ONLY project nr. 2017WR2LRS, which are gratefully acknowledged. GC also thanks the Alexander von Humboldt-Stiftung/ Foundation (Humboldt Research Award).

ACKNOWLEDGMENT

This manuscript summarizes concepts and discussion emerged from the large EU initiative SUNERGY (fossil-free fuels and chemicals for a climate-neutral Europe; www.sunergy-initiative.eu) which is acknowledged.

ABBREVIATIONS

$CO_2RR$ $CO_2$ reduction reaction; EJ exaJoule; EO ethylene oxide; EU European Union; FF fossil fuel; GDE gas diffusion electrode; GG greenhouse gas; NADE natural air diffusion electrode; NRR $N_2$ reduction reaction; NZE net zero emissions (of GG); OER oxygen evolution reaction; PEC photoelectrocatalytic; PEF PolyEthylene Furanoate; PEM Polymer electrolyte membrane; PET Polyethylene terephthalate; PINO Phthalimide N-Oxyl; PO propylene oxide; PtX Power-to-X; PV photovoltaic; RES Renewable energy sources; R&D Research and development; SOEC solid oxide electrolyzer cell; TEMPO Tetramethylpiperidine N-Oxyl; TWh terawatts per hour.

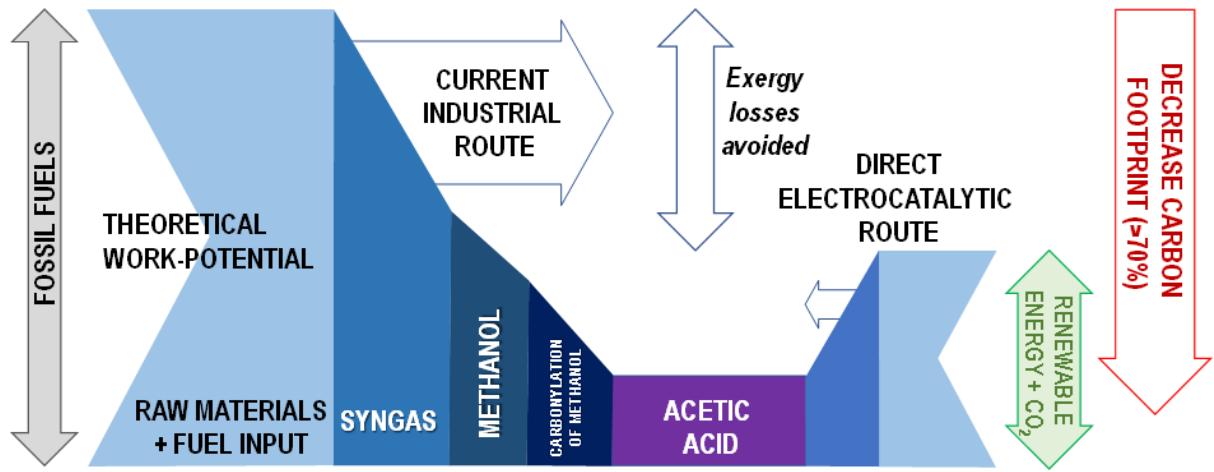

**Figure 1.** Grassmann-type diagram of indicative comparison of exergy in the multistep process to produce acetic acid via the conventional process starting from fossil fuels and the direct electrocatalytic route of $CO_2$ conversion to acetic acid with in-situ water electrolysis. Adapted with permission from Centi.[55]

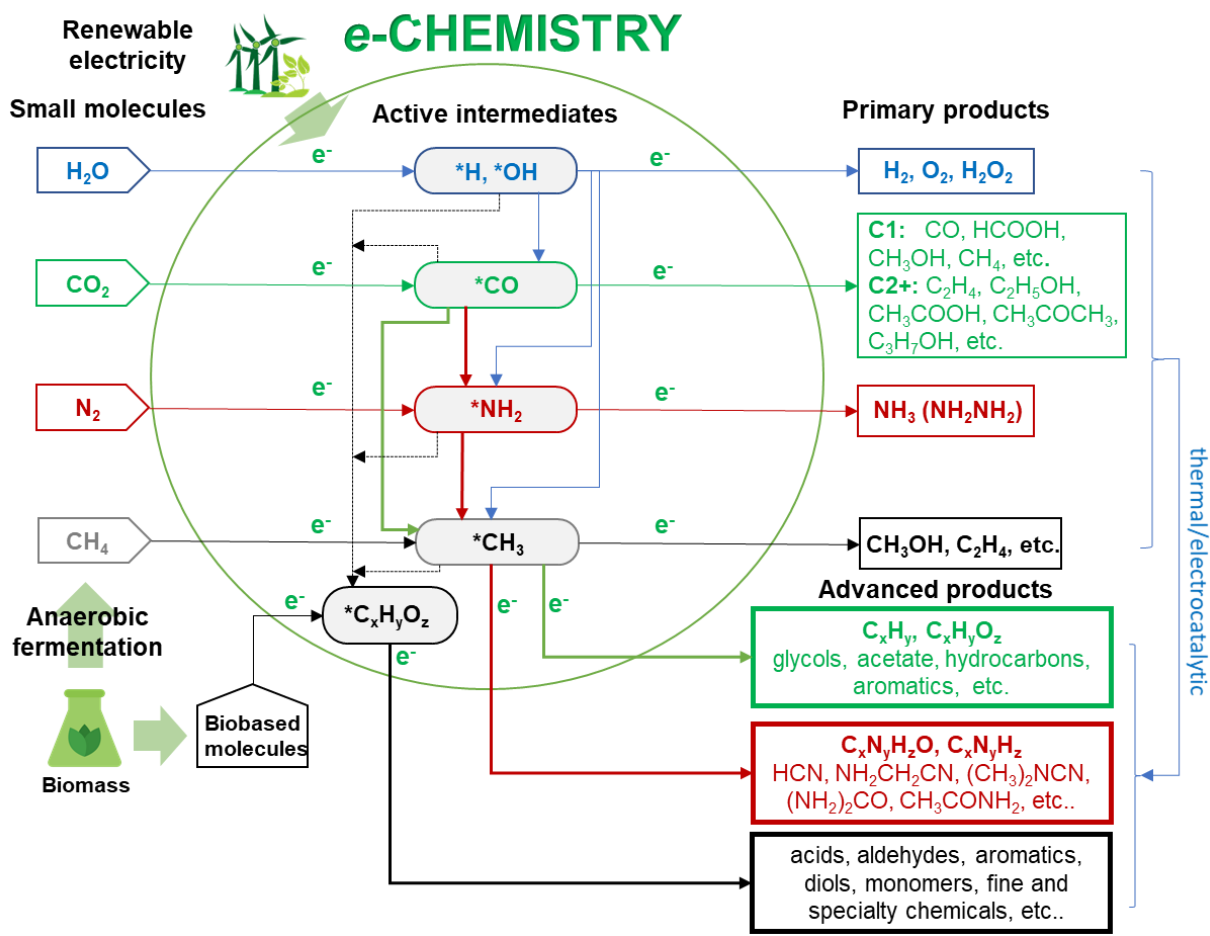

**Figure 2.** A new framework of electrocatalytically based reactions to develop a new *e*-chemistry alternative to that based on fossil fuels (petrochemistry). Adapted with permission from Perathoner[109] as full re-elaboration of the original concept presented by Tang et al.[69]

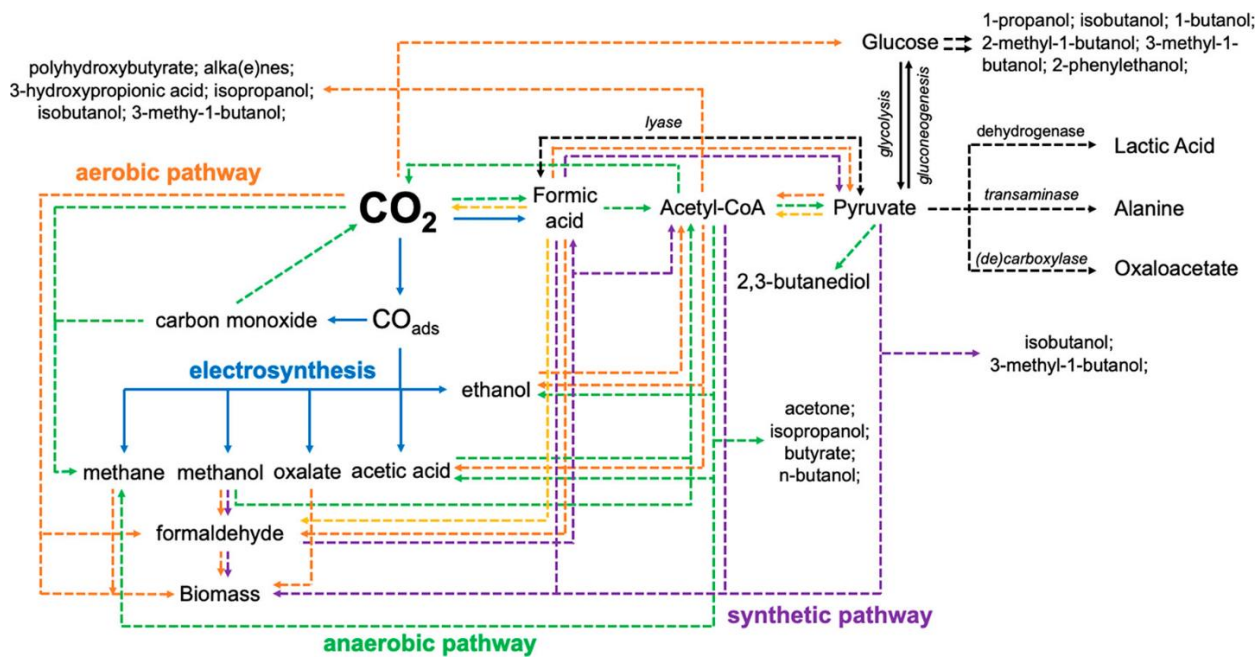

**Figure 3.** Integrated connectivity map of hybrid electro- and biosynthesis catalytic pathways in $CO_2$ conversion. Reproduced with permission from Atanassov and coworkers.[123]